\documentclass{elsart}
\usepackage{amssymb}
\usepackage[dvips]{epsfig}
\def \Emiss {\not\!\!E}
\newcommand{\gev}  {\mbox{\,GeV}}

\newcommand{\evcc} {\mbox{\,eV$/c^2$}}
\newcommand{\kevcc}{\mbox{\,keV$/c^2$}}
\newcommand{\gevcc}{\mbox{\,GeV$/c^2$}}
\newcommand{\tevcc}{\mbox{\,TeV$/c^2$}}
\newcommand{\ee}   {\mbox{e$^{+}$e$^{-}$}}
\newcommand{\grav} {\ensuremath{\mathrm{\tilde{G}}}}
\newcommand{\neu}  {\ensuremath{\tilde{\chi}_1^0}}
\newcommand{\slep} {\ensuremath{\tilde{\ell}}}
\newcommand{\sel}  {\ensuremath{\mathrm{\tilde{e}}}}
\newcommand{\smu}  {\ensuremath{\tilde{\mu}}}
\newcommand{\stau} {\ensuremath{\tilde{\tau}}}
\newcommand{\MP}    {\ensuremath{\mathrm{M_{P}}}}
\newcommand{\Mm}    {\ensuremath{\mathrm{M_{m}}}}
\newcommand{\tanb}  {\ensuremath{\tan\beta}}
\newcommand{\N}     {\ensuremath{\mathrm{N_{5}}}}
\newcommand{\rootF} {\ensuremath{\mathrm{\sqrt{F}}}}
\newcommand{\roots} {\ensuremath{\sqrt{s}}}
\begin{document}
\begin{frontmatter} 
\title{Searches for gauge mediated SUSY breaking at LEP}
\author{Ar\'an Garc\'\i a-Bellido}
\address{Royal Holloway University of London\\ Egham, Surrey TW20 0EX,
United Kingdom}
%\ead{Aran.Garcia-Bellido@cern.ch}
\begin{abstract}
The four LEP collaborations have performed searches for supersymmetric
particles in light gravitino scenarios, when neutralinos and sleptons are
produced and may present measurable decay lengths. The highest energy data
has been analysed including centre-of-mass energies up to 209$\gev$. No
evidence for such particles is found, but preliminary limits from the
combination allow to exclude at 95\% confidence level neutralino masses up
to 97$\gevcc$ if neutralinos decay promptly and stau masses up to 
86.9$\gevcc$ for all stau lifetimes. 
The interpretation of these results in general models is studied to set
limits on the parameters of the theory. 
\end{abstract}
\end{frontmatter}

\section{Introduction}
The fact that supersymmetry (SUSY) is a broken symmetry makes searches for the
predicted super-partners very rich phenomenologically. 
If SUSY is broken 
spontaneously at some very high energy scale, several models exist that
predict how this breaking is communicated to the MSSM particles. 
If gravitational interactions are responsible for the MSSM spectrum, 
the neutralino is most probably the lightest supersymmetric particle (LSP) and
these models are usually referred to as SUGRA~\cite{sugra}.
Another possibility is that
some new very heavy particles, called messengers, couple radiatively to the
MSSM particles via gauge interactions. This model is the subject of the
present note and is called gauge mediated supersymmetry breaking or
GMSB~\cite{gmsb_theory}. The main difference between gravity and gauge
mediated SUSY breaking models arises from the higher energy scale of SUSY
breaking in the former. This will 1) introduce possible flavour changing
neutral currents (FCNC), severely constrained at the electroweak
scale; and 2) make the gravitino, 
the massive spin-$\frac{3}{2}$ partner of the graviton consequence of the
super-Higgs mechanism, as heavy as the rest of SUSY particles with a mass
of at least some $\mathcal{O}(100)\gevcc$. 
In gravity mediated SUSY breaking models the MSSM masses are
linked directly with the Planck mass $\MP\sim 10^{19}\gevcc$, 
where gravity is assumed to become strong. Alternatively, 
if a new
messenger sector is introduced like in GMSB models with common mass $\Mm$
as low as $10^{4}\gevcc$ and clearly much lower than $\MP$, then: 
\renewcommand{\labelenumi}{\theenumi .}
\vspace{-0.3cm}
\begin{enumerate}
\item FCNC effects are naturally suppressed because the SM Yukawa couplings are
already present when MSSM mass generation takes place and gauge
interactions are flavour blind. Thus squark and slepton mass differences
are very small.   
\item The gravitino becomes very light:
$m_{\grav}\lesssim\mathcal{O}(1)\kevcc$~\cite{cosmo}, and is the LSP. But
it will not only couple with gravitational strength
to other SUSY particles. Its longitudinal components allow it to interact
with the next-to-lightest supersymmetric particle (NLSP) via the SM gauge
couplings. 
\end{enumerate}

Six new parameters, in addition to those of the SM, are needed to
completely specify the MSSM spectrum and phenomenology of minimal GMSB
models. These are: $\rootF$, the energy scale of SUSY breaking in the
hidden sector; $\Mm$, the common mass of the messenger particles; $\N$, the
number of SU(5) families of messenger supermultiplets; $\Lambda$, the mass
scale parameter responsible for the MSSM masses at $\Mm$; $\tanb$, the
ratio of the vacuum expectation values of the two Higgs doublets; and
sign($\mu$), where $\mu$ is the Higgs mixing-mass parameter. 

GMSB signatures at colliders crucially depend on the nature of the
NLSP. In general, for $\N=1$ the neutralino is the NLSP and decays to
$\gamma\grav$, while for $\N\ge2$
the lightest stau is predominantly the NLSP and decays into $\tau\grav$. 
In GMSB models, the selectron
and smuon masses are equal and the stau may become lighter due to the
possible large mixing in the third family. But depending on the scale of
SUSY breaking $\rootF$, or equivalently on the gravitino mass $m_{\grav}$,
the NLSP may be long-lived: 
\begin{center}
\[ 
\lambda_{\rm NLSP} = c\tau_{\rm NLSP}\gamma\beta = 
\frac{0.01}{\kappa_{\gamma}}\left( \frac{100\gevcc}{m_{\rm NLSP}}\right)^{-5}
\left( \frac{m_{\grav}}{2.4\evcc} \right)^2
\sqrt{\frac{E^2}{m_{\rm NLSP}^2}-1}~{\rm cm}
\]
\end{center}
where $\kappa_{\gamma}$ is the photino content of the neutralino or one for
a slepton NLSP. Since the gravitino mass can range from $\sim$$10^{-2}$ to
$10^{5}\evcc$ this implies that the NLSP may decay immediately after
production at LEP, somewhere inside the detector or be stable for searches
purposes. There are thus many different topologies to be considered as a
function of the NLSP type and its possible decay length. 

This report is organised as follows: topologies arising from
neutralino pair production in the neutralino NLSP scenario are described
first. The results from slepton NLSP direct pair-production searches are
reported next in Sec.~\ref{slep}. 
Cascade decays of neutralino to slepton NLSPs and other
multilepton final states are described in Sec.~\ref{multilep}. 
In each case, the different searches devised to cover all possible
lifetimes of the NLSP will be detailed. 
Finally, the interpretation of the results and outlook will be presented in
Secs.~\ref{scan} and~\ref{concl} respectively. 
All limits are given at 95\% confidence level. 

\section{Photon(s) and {\boldmath$\Emiss$} signatures}
\begin{figure}[tb]
\begin{center}
\includegraphics[width=0.35\textwidth]{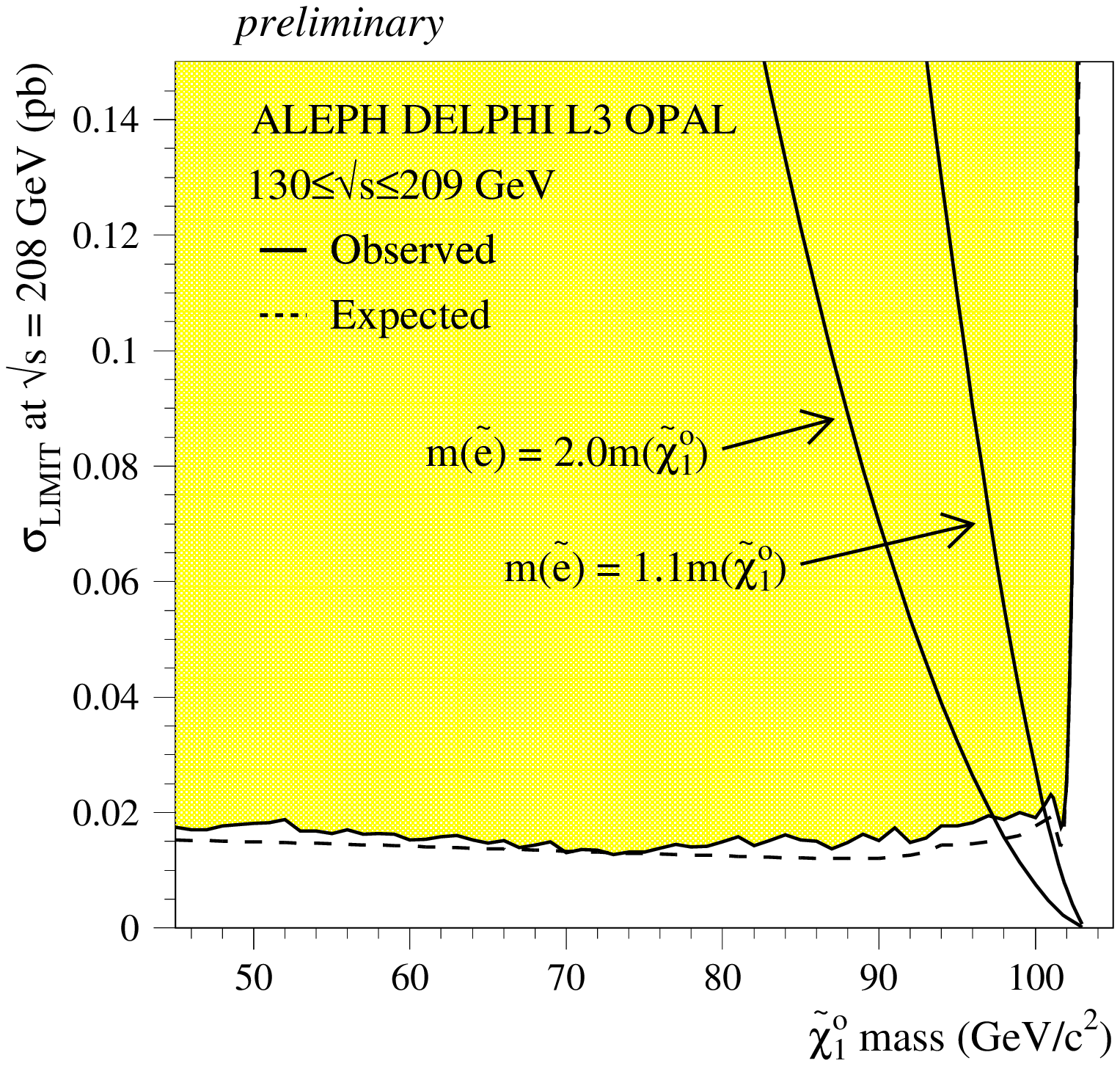}
\includegraphics[width=0.35\textwidth]{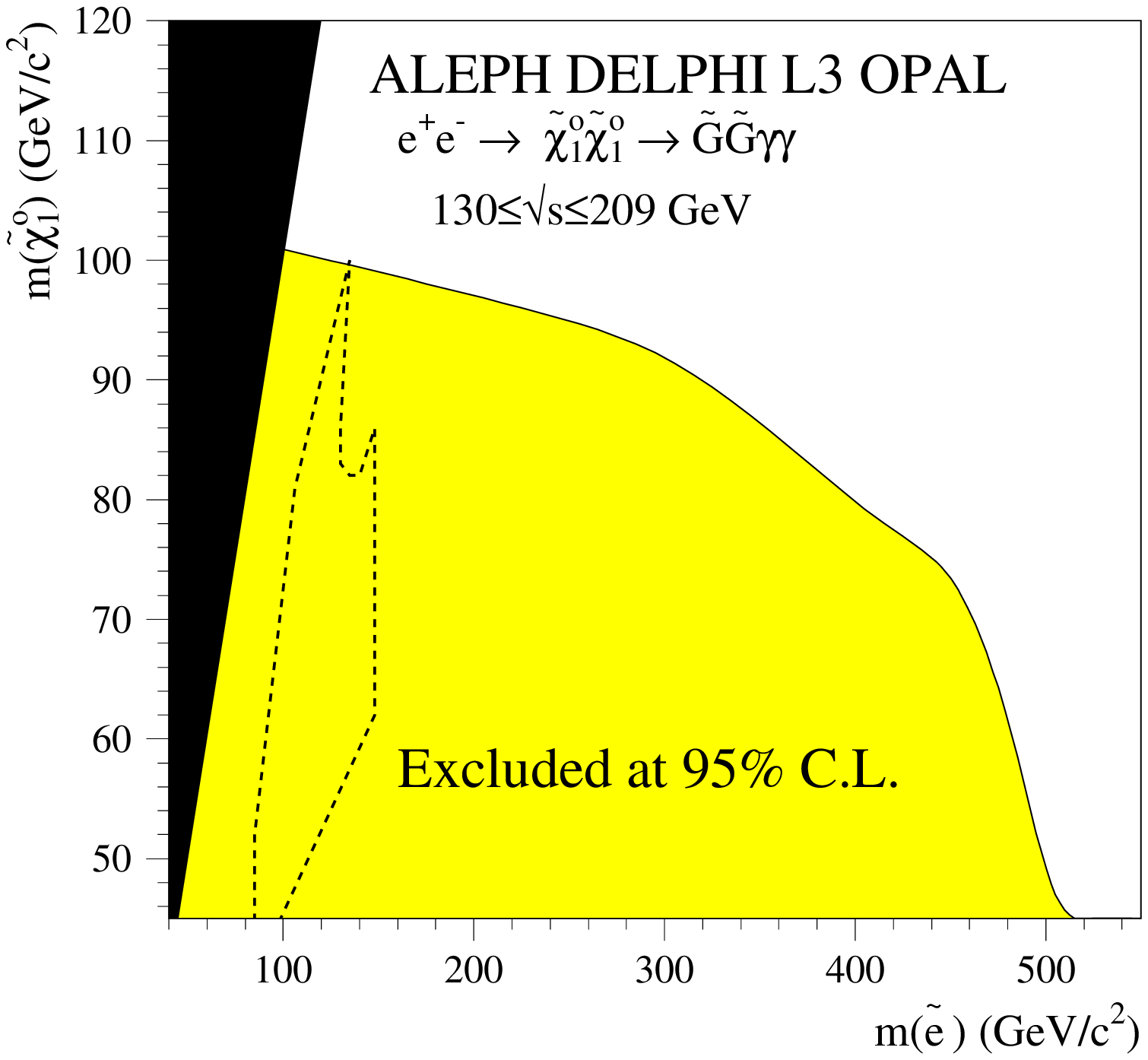}
\caption{\label{fig:twophot}{\small Left: cross section lower limit versus
neutralino mass from the combined two-photon searches. The production cross
sections are calculated in two scenarios. Right: excluded area in the
selectron-neutralino  mass plane assuming a purely bino neutralino. The
dashed overlayed area is the region consitent with the GMSB
interpretation of the ee$\gamma\gamma+\Emiss_{\rm
T}$ CDF event~\cite{Ambrosanio:1996zr}, now completely excluded.}} 
\end{center}
\end{figure}

Signatures consisting of two energetic photons and missing energy may arise
in the neutralino NLSP scenario if a pair of neutralinos are pair-produced
and decay promptly into $\neu\to\gamma\grav$. The main background source
for this topology is $\ee\to\nu\bar{\nu}$, 
from diagrams with $s$-channel Z exchange or $t$-channel W exchange and
one or two initial state radiation (ISR) photons. 
By imposing a threshold cut on the energy of the least energetic photon
the SM background can be effectively reduced. A preliminary combination of
the results from the ALEPH, DELPHI and OPAL collaborations exists 
with data from 192 to 208$\gev$ and yields no excess with respect to the SM
expectation~\cite{LSWGweb}. The individual results can be found in
Refs.~\cite{aleph_phot,delphi_phot,l3_mega,opal_gmsb}. 
Figure~\ref{fig:twophot} shows the interpretation of these results assuming
a purely bino neutralino and mass degeneracy between $\sel_{\rm L}$ and
$\sel_{\rm R}$. Neutralino pair-production cross sections above 0.02\,pb 
are excluded at 95\% C.L. 
Since the cross section depends on the mass of the
exchanged selectron, the lower limit on the neutralino mass can be
conservatively estimated to be $97\gevcc$ if $m_{\sel}=2\cdot m_{\neu}$,
improving for larger $\neu-\sel$ mass differences. 

If the gravitino mass lies in the range between a few $\evcc$ and a few
hundred $\evcc$, the neutralino NLSP will decay somewhere inside the
detector. 
The probability for only one of the pair-produced neutralinos to
decay before the electromagnetic calorimeter is greater than for both of
them~\cite{gmsb_theory}, thus the expected final topology consists in a single
non-pointing photon. ALEPH~\cite{aleph_phot} and DELPHI~\cite{delphi_phot}
have searched for single photons with a minimum impact parameter of 40\,cm
and have found 2/1.0 and 5/3.0 candidates/background, 
respectively in their $\roots=189-209$ and $\roots=202-209$ data sets.  
The efficiency for this search reaches a 
maximum of around 10\% for decay lengths of 8\,m. This yields an approximate
upper limit of 0.4\,pb in the production cross section.

\section{Two leptons and {\boldmath$\Emiss$} signatures}
\label{slep}
We discuss next the pair-production of sleptons in the slepton NLSP
scenario. In general the $\stau_1$ will be the lighter slepton, but a
degenerate case is possible in which the three $\stau_1$, $\sel_{\rm R}$ and
$\smu_{\rm R}$ act as co-NLSP's.  Thus searches for pairs of sleptons
decaying into their SM counterpart and a gravitino were developed by all
four collaborations taking into account the possible lifetime of the
slepton. 
\subsection{Acoplanar leptons}
In the case of zero slepton lifetime two acoplanar leptons and missing
energy are expected. This is the same final state as the one arising in
SUGRA models if we take a massless neutralino. 
The existing limits at 95\% C.L. on the slepton masses derived from the 
combination~\cite{LSWGweb} for $m_{\neu}=0$ are: 
$m_{\sel} > 99.6 , m_{\smu} > 94.9$ and $m_{\stau} > 85.0\gevcc$.
\subsection{Kinks and large impact parameters}
When sleptons decay in the tracking chambers of the detectors they are
expected to produce kinks formed by their own track and the emerging
lepton. If they do not reach the tracking devices, events will present
tracks with large impact parameters.
In these searches the most important sources of background are hadronic
interactions in the material of the detector from K$_{\rm{s}}^0$ for the
large impact parameter topology and K$^{\pm}$
or $\pi^{\pm}$ decays in the kinks search. Both can be reduced by cuts on
the opening angle between the two segments and with energy vetos. 
ALEPH reports 1/1.1 candidates/background, DELPHI 9/7.4 and OPAL 1/1.1 for
selectrons and smuons and 7/4.4 for staus in their $\roots=189-209\gev$ data. 
\subsection{Heavy stable charged particles}
For large gravitino masses, above a few hundred $\evcc$, the sleptons decay
outside the detector. This topology is characterised by two back-to-back
highly ionising tracks which if produced at threshold may completely
saturate the chamber electronics by their high d$E$/d$x$. Very high
efficiencies can be attained in this search and no collaboration has
found any discrepancy with the expected background. The combination of
results from the four collaborations yields a lower limit on
$m_{\smu}$ of $97.5\gevcc$ for NLSP lifetimes greater than
$10^{-6}$\,s.  
\begin{figure}[tbp]
\begin{center}
\includegraphics[width=0.31\textwidth]{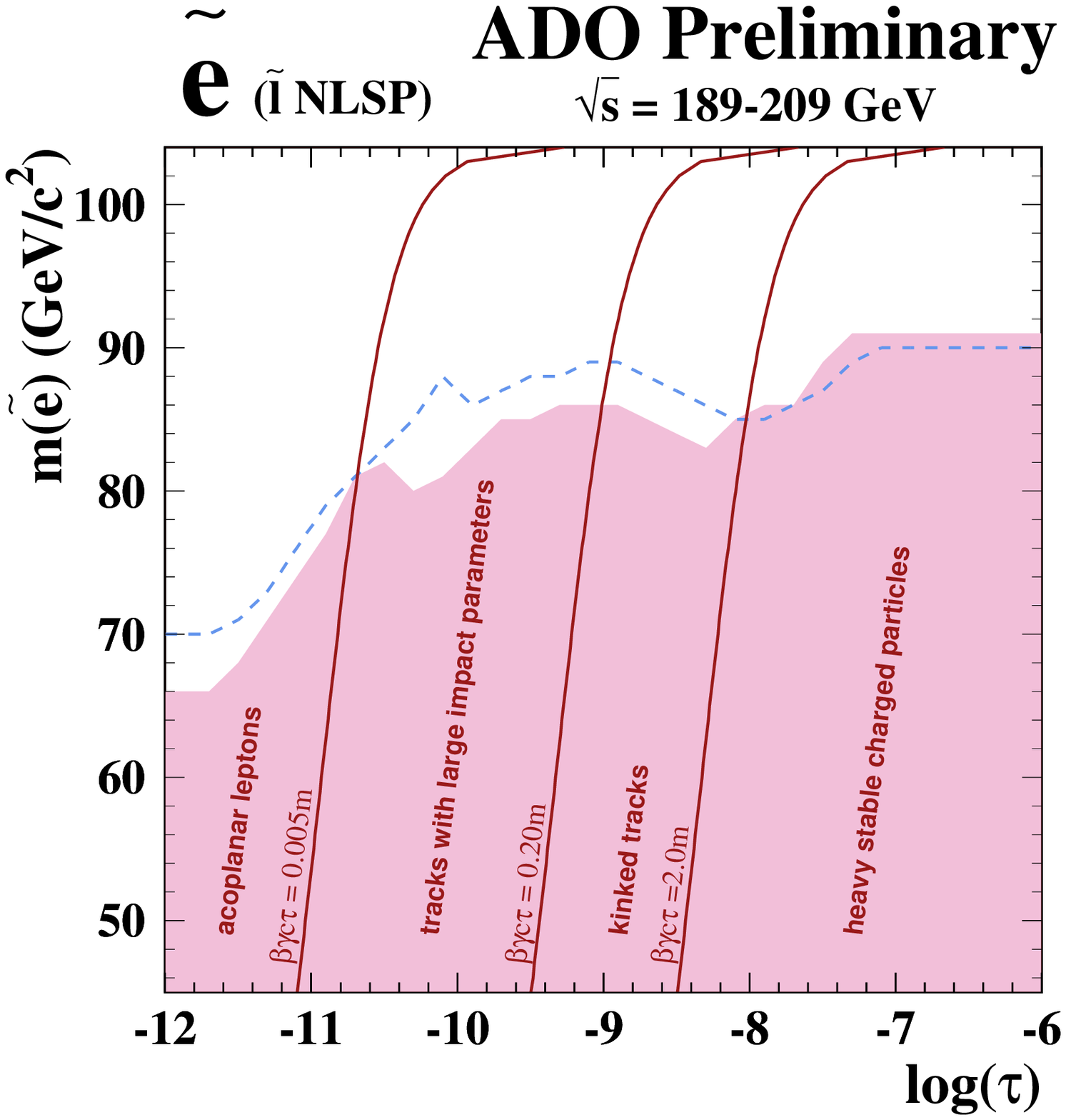}
\includegraphics[width=0.31\textwidth]{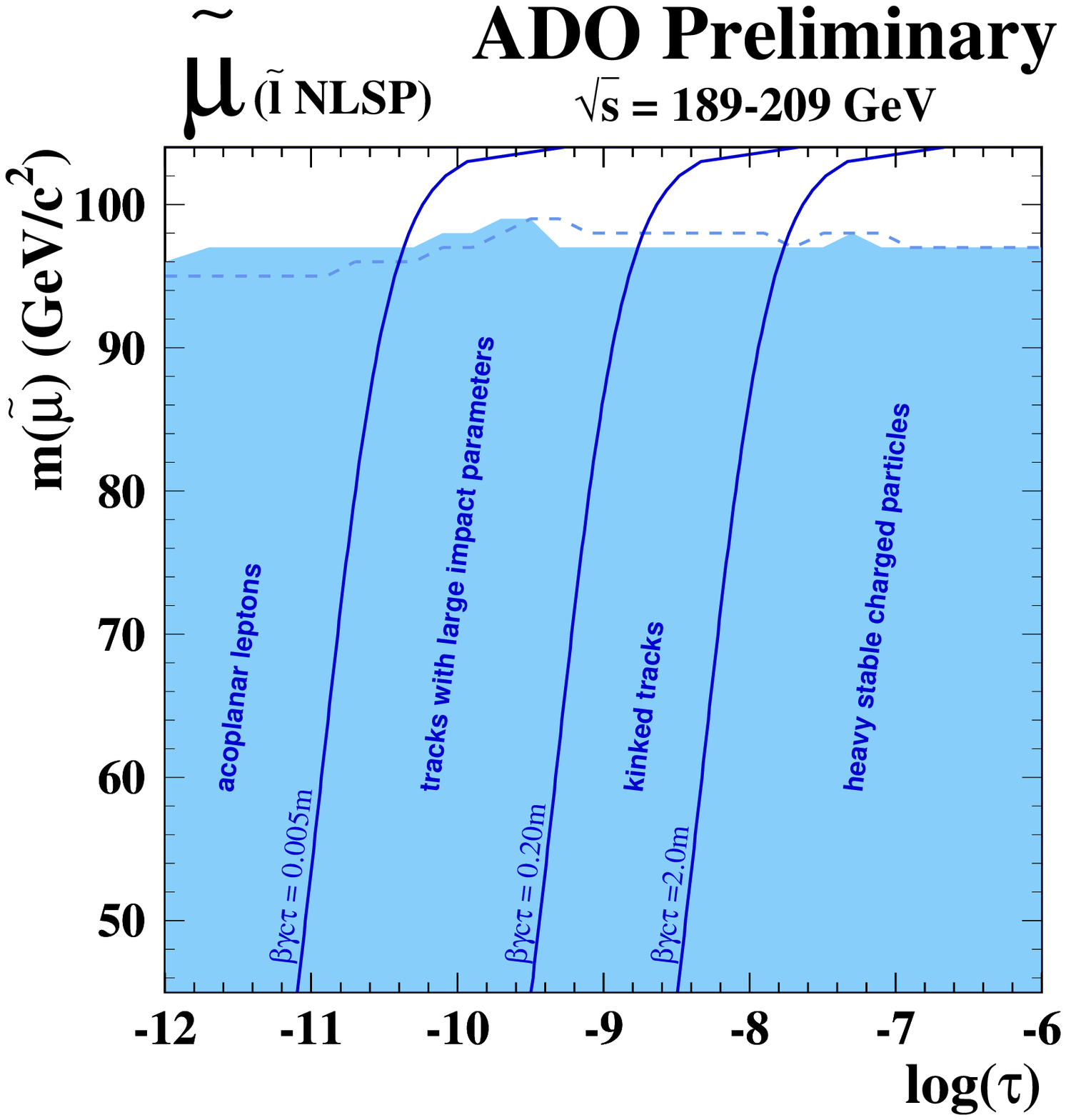}
\includegraphics[width=0.31\textwidth]{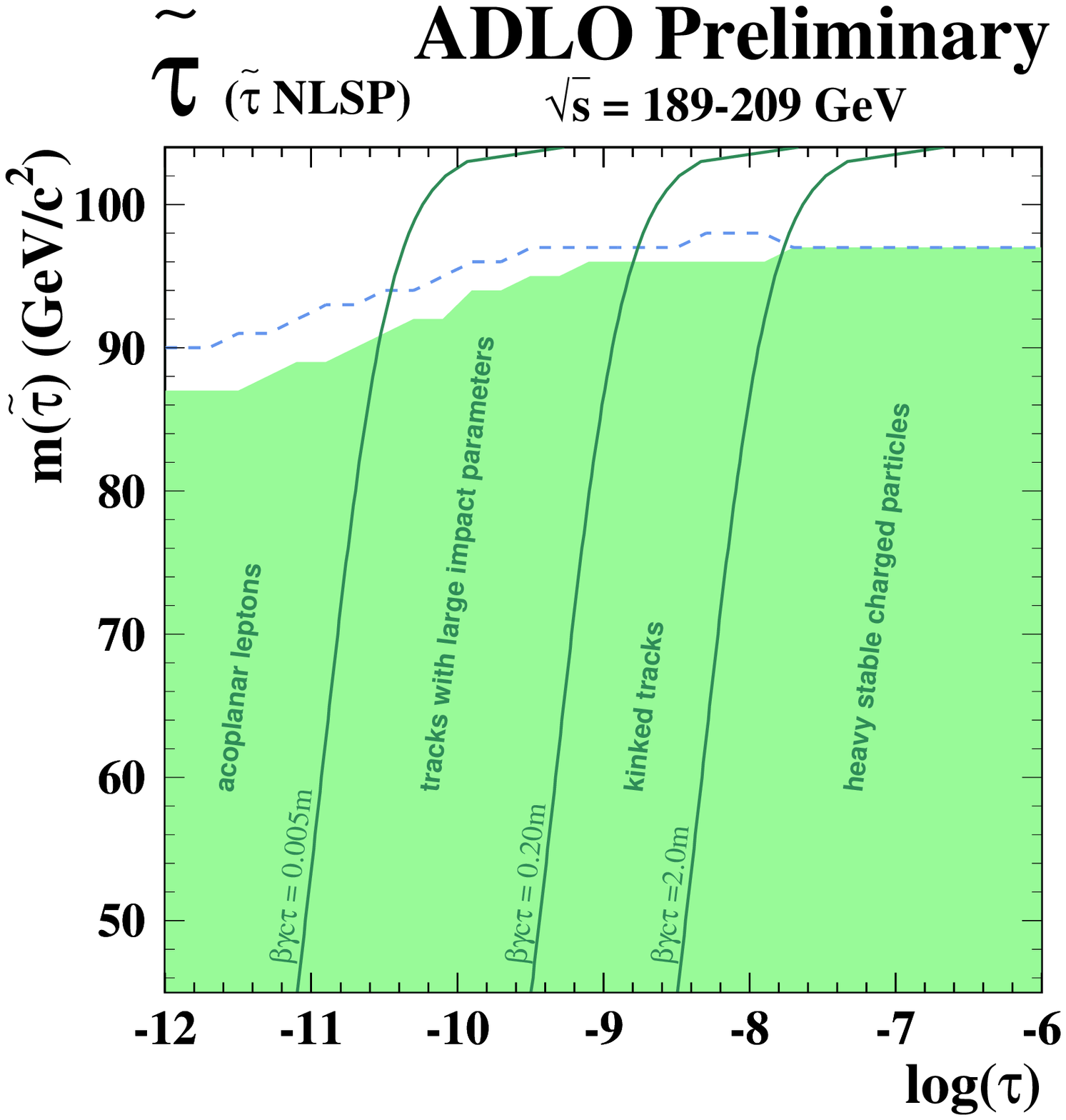}
\caption{\label{fig:slep}{\small  Combined exclusion on the selectron,
smuon and stau masses as a function of lifetime for sleptons co-NLSP's and
stau NLSP. The combined preliminary limits are set at: 
$m_{\sel} > 65.8 , m_{\smu} > 96.3$ and $m_{\stau} > 86.9\gevcc$, valid for
any lifetime and $\N\leq5$.}}
\end{center}
\end{figure}

\section{Multi-lepton and {\boldmath$\Emiss$} signatures}
\label{multilep}
If neutralinos are accessible at LEP in the slepton NLSP scenario, they
provide a clear advantadge over slepton direct production: a higher cross
section. Indeed, their production cross section may be even twice that of
sleptons, specially for very light selectrons. Searches have been developed
by ALEPH~\cite{aleph_gmsb}, DELPHI~\cite{delphi_gmsb} and
OPAL~\cite{opal_gmsb} for four leptons (two of them could be soft) and
missing energy in the final state. Only ALEPH and OPAL have also explored
the possibility of two of the tracks exhibiting large impact parameters or
kinks. The results are again negative. In the case of zero lifetime this
search helps to exclude an area in parameter space that searches for
direct slepton production do not reach. 
In addition to neutralino production in the slepton NLSP scenario, OPAL and
DELPHI have looked for charginos $\tilde{\chi}^+_1\to\slep^+_{\rm R}\nu$
and OPAL and ALEPH for selectrons or smuons in the stau NLSP case:
$l^+_{\rm R}\to l^+\neu\to l^+\tau\stau_1$, where $l$ = e, $\mu$. 
No evidence for any of the above has been found 
and limits have been set on the mass of charginos (100$\gevcc$ for any
lifetime by DELPHI) and sleptons (94$\gevcc$ for selectrons and zero
lifetime by ALEPH) in the $\stau_1$ NLSP scenario.  
\section{Interpretation in mGMSB}
\begin{figure}[tbp]
\begin{center}
\includegraphics[width=0.31\textwidth]{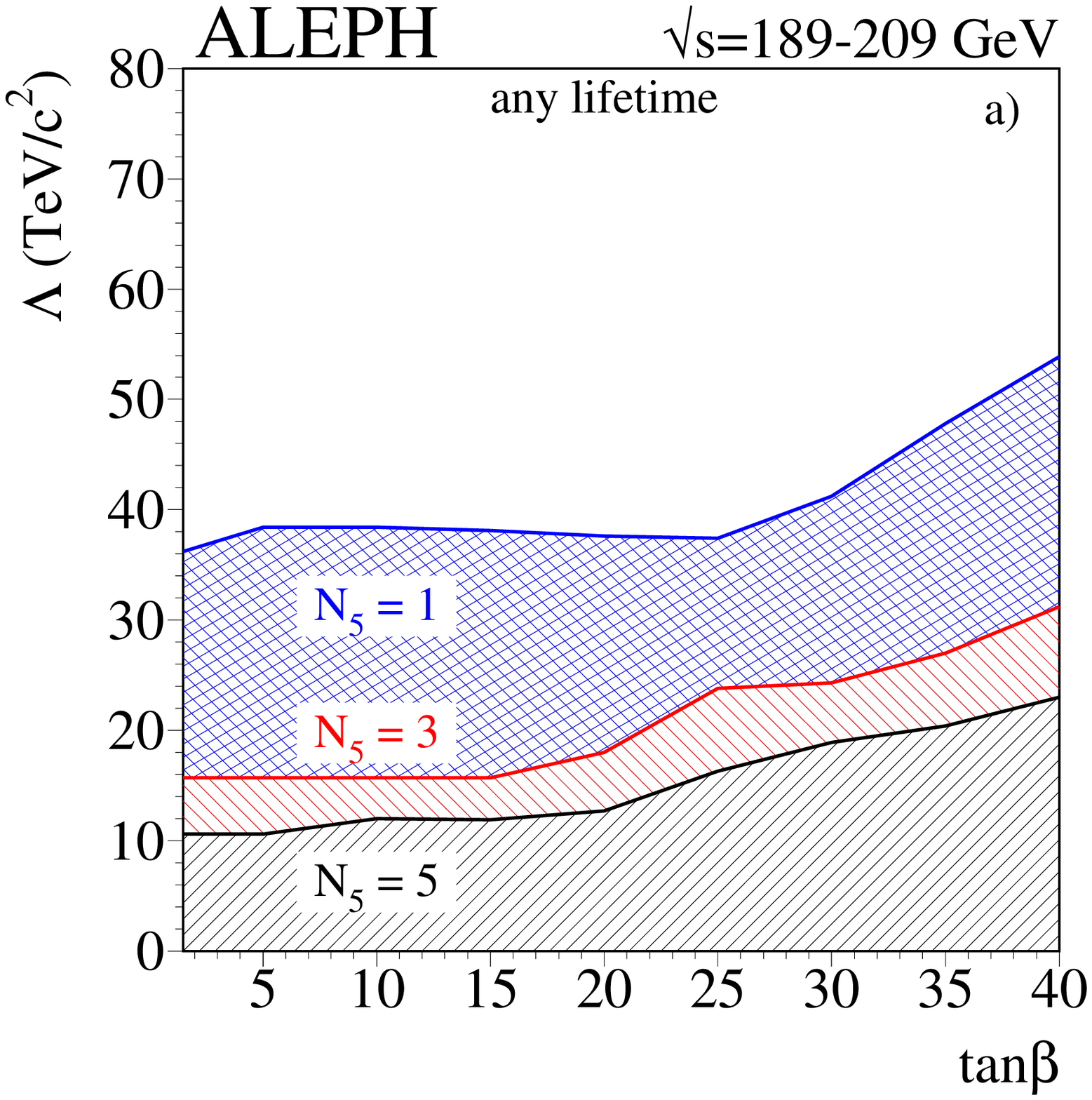}
\includegraphics[width=0.31\textwidth]{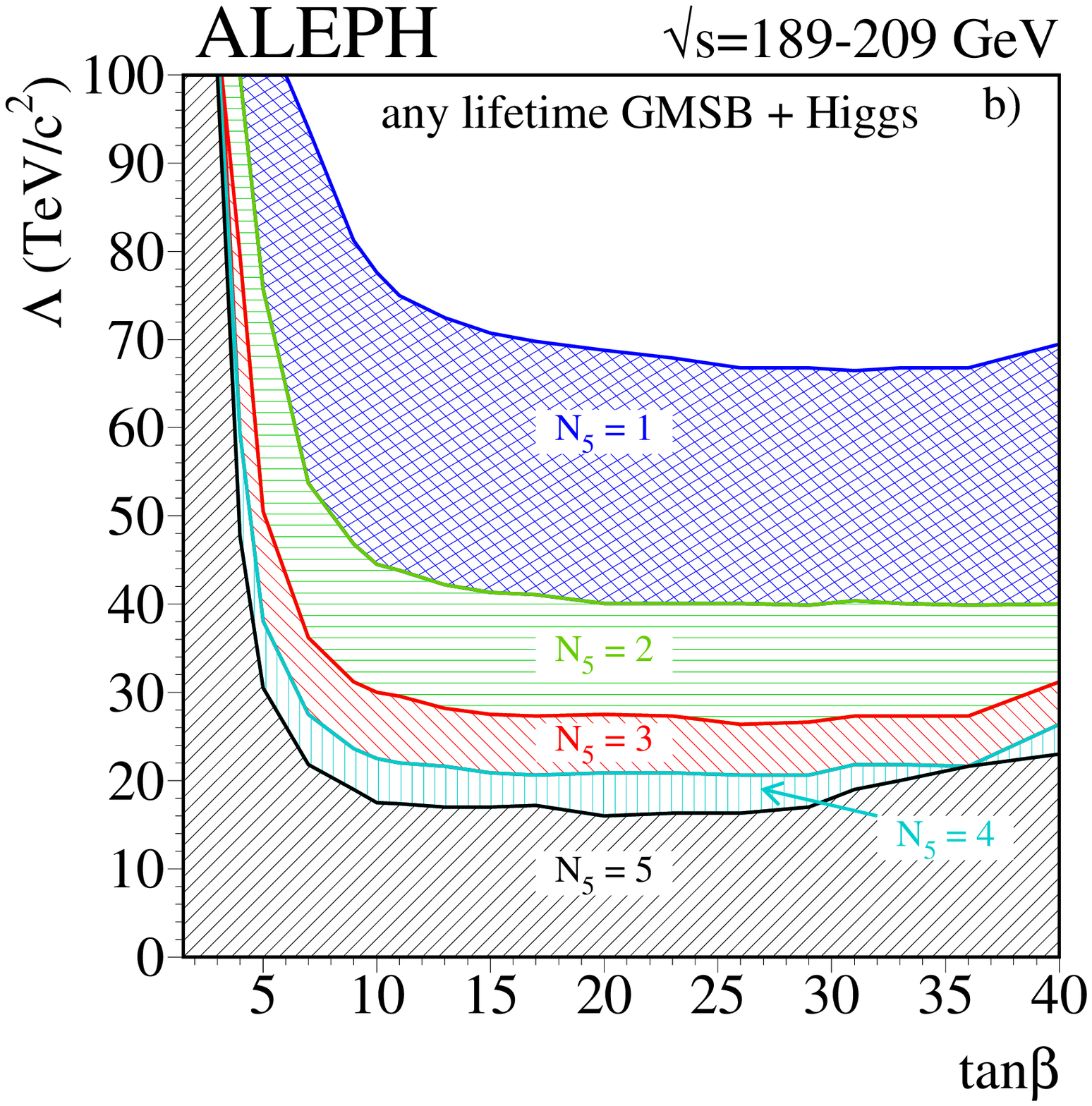}
\includegraphics[width=0.31\textwidth]{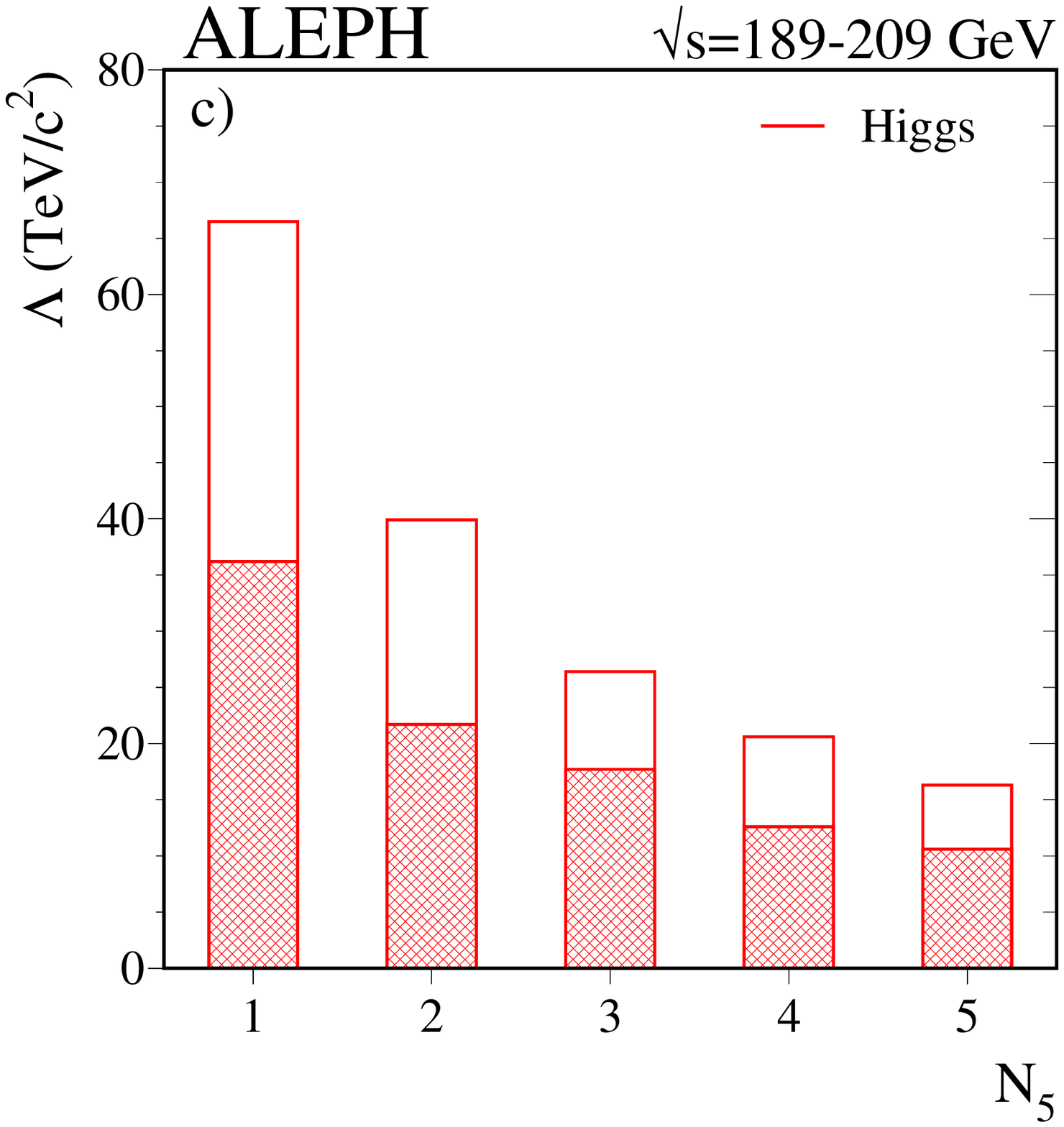}
\caption{\label{fig:scan}{\small Excluded values of the mass scale
parameter $\Lambda$ as a function of $\tanb$ for different values of $\N$,
as set by ALEPH searches for (a) GMSB topologies and (b) neutral Higgs
bosons. The minimum allowed value of $\Lambda$ as a function of $\N$ (c) if
GMSB only (shaded) or Higgs only (unshaded) searches are included in the
exclusion procedure. From Ref.~\cite{aleph_gmsb}.}}
\end{center}
\end{figure}

By using the upper limits on the cross sections, the searches results
contrain the available parameter space in minimal GMSB models. ALEPH,
DELPHI and OPAL have performed scans over the six variables that determine
GMSB phenomenology, $\rootF$, $\Mm$, $\Lambda$, $\N$, $\tanb$ and
sign($\mu$) to asses the impact of the different searches and set limits on
these parameters. An example from ALEPH is shown in Fig.~\ref{fig:scan}, 
where the ($\Lambda,\tanb$) plane is shown with the excluded areas. 
This analysis places a lower limit on the mass scale parameter $\Lambda$ of
10$\tevcc$ for any lifetime and $\N\leq5$, which is increased to 16$\tevcc$
if Higgs boson searches are included, as can be seen in
Fig.~\ref{fig:scan}. OPAL results (without Higgs limits) are able exclude
$\Lambda>40,27,21,17,15\tevcc$ for $\N=1,2,3,4,5$ respectively, 
taking all NLSP lifetimes into account in a similar scan to
the one by ALEPH. DELPHI reports a limit of $\Lambda>17.5\tevcc$ for
$\N\leq4$ and negligible NLSP lifetimes. 
\label{scan}

\section{GMSB after LEP}
\label{concl}
To conclude, the LEP SUSY working group has performed the combination of
available results in the most important searches within the GMSB
framework. They have been presented here, including some other results that
the collaborations have performed to completely cover most of the GMSB
parameter space. No hint for any of these topologies has been found. 
Neutralino and stau masses have been excluded up to
54 and 87$\gevcc$ in the neutralino and stau NLSP scenarios
respectively and independently of lifetime. 
This can be translated into a lower limit on the mass scale parameter
$\Lambda > 10\tevcc$ for $\N\leq5$. 
The quest for SUSY is now open to the Tevatron RunII. 
Chargino masses up to 290 (160)$\gevcc$ could be discovered by 2004 with
2\,fb$^{-1}$ in the $\neu$ ($\stau_1$) NLSP scenario with short (long)
lifetimes, as described in Ref.~\cite{Culbertson:2000am}. 
ALEPH alone for example has excluded chargino masses of
160 (100)$\gevcc$ for short (long) NLSP lifetimes.  
\section*{Acknowledgements}
\vspace{-0.5cm}
I am very grateful to Christoph Rembser and Rob McPherson for providing 
their help (and results!) for this talk.

\end{document}